\documentclass[12pt]{article}
\usepackage{amsmath}
\usepackage{amssymb}

\input{tcilatex}
\begin{document}

\bigskip

\bigskip \ 

\bigskip \ 

\begin{center}
\textbf{QUBITS AND ORIENTED MATROIDS IN}

\textbf{\ }

\textbf{FOUR TIME AND FOUR SPACE DIMENSIONS}

\textbf{\ }

\smallskip \ 

J. A. Nieto\footnote{%
niet@uas.edu.mx, janieto1@asu.edu}

\smallskip

\textit{Facultad de Ciencias F\'{\i}sico-Matem\'{a}ticas de la Universidad
Aut\'{o}noma de Sinaloa, 80010, Culiac\'{a}n Sinaloa, M\'{e}xico}

and

\textit{Mathematical, Computational \& Modeling Sciences Center, Arizona
State University, PO Box 871904, Tempe, AZ 85287, USA}

\smallskip \ 

\bigskip \ 

\bigskip \ 

Abstract
\end{center}

We establish a connection between 4-rebits (real qubits) and the Nambu-Goto
action with target `spacetime' of four time and four space dimensions
((4+4)-dimensions)). We motivate the subject with three observations. The
first one is that a 4-rebit contains exactly the same number of degree of
freedom as a complex 3-qubit and therefore 4-rebits are special in the sense
of division algebras. Secondly, the (4+4)-dimensions can be splitted as
(4+4)=(3+1)+(1+3) and therefore they are connected with an ordinary
(1+3)-spacetime and with changed signature (3+1)-spacetime. Finally, we show
how geometric aspects of 4-rebits can be related to the chirotope concept of
oriented matroid theory.

\bigskip \ 

\bigskip \ 

\bigskip \ 


Pacs numbers: 04.60.-m, 04.65.+e, 11.15.-q, 11.30.Ly

December, 2012

\newpage

Recently, through the identification of the coordinates $x^{\mu }$ of a
bosonic string, in target space of $(2+2)$-signature, with a $2\times 2$
matrix $x^{ab}$, Duff [1] was able to discover new hidden discrete
symmetries of the Nambu-Goto action [2]-[3]. It turns out that the key
mathematical tool in this development is the Cayley hyperdeterminant $Det(b)$%
\ [4] of the hypermatrix $b_{a}^{~~bc}=\partial _{a}x^{bc}$. A striking
result is that $Det(b)$ can also be associated with the four electric
charges and four magnetic charges of a STU black hole in four dimensional
string theory [5]. Even more surprising is the fact that $Det(b)$ makes also
its appearance in quantum information theory by identifying $b_{a}^{~~bc}$
with a complex $3$-qubit system $a_{a}^{~~bc}$ [6]. These coincidences,
among others, have increased the interest on the qubit/black hole
correspondence [7].

It has been shown [8] that a straightforward generalization of the above
Duff's formalism, concerning the Nambu-Goto action, can be applied to a
target space of $(5+5)$-signature, but not to a space of $(4+4)$-signature.
But, since in principle, the $(5+5)$-signature may be associated with a $5$%
-qubit and the $(4+4)$-signature with a $4$-qubit this is equivalent to say
that the Nambu-Goto action exhibit discrete symmetries for a $5$-qubit
system, but not for a $4$-qubit system.

On the other hand, in quantum information theory it does not seem to be any
particular reason for avoiding unnormalized $4$-qubits. In fact, a $4$-qubit
is just one possibility out of the complete set of $N$-qubit systems. It
turns out that, in a particular subclass of $N$-qubit entanglement, the
Hilbert space can be broken into the form $C^{2^{N}}=C^{L}\otimes C^{l}$,
with $L=2^{N-1}$ and $l=2$. Such a partition it allows a geometric
interpretation in terms of the complex Grassmannian variety $Gr(L,l)$ of $2$%
-planes in $C^{L}$ via the Pl\"{u}cker embedding. In this case, the Pl\"{u}%
cker coordinates of Grassmannians $Gr(L,l)$ are natural invariants of the
theory. It turns out that in this scenario the complex $3$-qubit, $4$-qubit
and $5$-qubit admit a geometric interpretation in terms of the complex
Grassmannians $Gr(4,2)$, $Gr(8,2)$ and $Gr(16,2)$, respectively (see Refs
[9] and [10] for details).

Of course, in this context, it has been mentioned in Ref. [11], and proved
in Refs. [12] and [13], that for normalized qubits the complex $1$-qubit, $2$%
-qubit and $3$-qubit are deeply related to division algebras via the Hopf
maps, $S^{3}\overset{S^{1}}{\longrightarrow }S^{2}$, $S^{7}\overset{S^{3}}{%
\longrightarrow }S^{4}$ and $S^{15}\overset{S^{7}}{\longrightarrow }S^{8}$,
respectively. It seems that there does not exist a Hopf map for higher $N$%
-qubit states. So, from the perspective of Hopf maps, and therefore of
division algebras, one arrives to the conclusion that $1$-qubit, $2$-qubit
and $3$-qubit are more special than higher dimensional qubits (see Refs.
[11]-[13] for details).

How can we make sense out of these different scenarios in connection with a $%
4$-qubit system? Before we try to answer this question, let us think in a $3$%
-qubit/black hole correspondence. In this case the symmetry of a extremal
STU black hole model is $SL(2,R)^{\otimes 3}$. However in the case of a
complex qubit system the symmetry group is $SL(2,C)^{\otimes 3}$. So, the
problem is equivalent to an embedding of a real $3$-qubit ($3$-rebit, see
Ref. [14] for definition of $N$-rebits) relevant in STU black holes into
complex $3$-qubit in complex geometry. It has been shown [9] that this kind
of embedding is not trivial and in fact requires the mathematical tools of
fiber bundles with Gramannian variety as a base space. It has been compared
[10] this mechanism with the analogue situation described in twistor theory
when one pass from real to complex Minkowski space (see also Refs.
[15]-[17]).

Apart from these embeddings one may gain some insight on the above subject
if one simple counts the number of degrees of freedom corresponding to the
complex $3$-qubit and $4$-qubit and compare them with the corresponding real
qubits, $3$-rebit, $4$-rebit. Consider the general complex state $\mid \psi
>\in C^{2^{N}},$

\begin{equation}
\mid \psi >=\dsum
\limits_{a_{1},a_{2},...,a_{N}=0}^{1}a_{a_{1}a_{2}...a_{N}}\mid
a_{1}a_{2}...a_{N}>,  \tag{1}
\end{equation}%
where the states $\mid a_{1}a_{2}...a_{N}>=\mid a_{1}>\otimes \mid
a_{2}>...\otimes \mid a_{N}>$ correspond to a standard basis of the $N$%
-qubit. For a $3$-qubit (1) becomes

\begin{equation}
\mid \psi >=\dsum \limits_{a_{1},a_{2},a_{3}=0}^{1}a_{a_{1}a_{2}a_{3}}\mid
a_{1}a_{2}a_{3}>,  \tag{2}
\end{equation}%
while for $4$-qubit one has

\begin{equation}
\mid \psi >=\dsum
\limits_{a_{1},a_{2},a_{3},a_{4}=0}^{1}a_{a_{1}a_{2}a_{3}a_{4}}\mid
a_{1}a_{2}a_{3}a_{4}>.  \tag{3}
\end{equation}%
One observes that $a_{a_{1}a_{2}a_{3}}$ has $8$ complex degrees of freedom,
that is $16$ real degrees of freedom, while $a_{a_{1}a_{2}a_{3}a_{4}}$
contains $16$ complex degrees of freedom, that is $32$ real degrees of
freedom. Let us denote $N$-rebit system (real $N$-qubit ) by $%
b_{a_{1}a_{2}...a_{N}}$. So we shall denote the corresponding $3$-rebit, $4$%
-rebit by $b_{a_{1}a_{2}a_{3}}$ and $b_{a_{1}a_{2}a_{3}a_{4}}$,
respectively. One observes that $b_{a_{1}a_{2}a_{3}}$ has $8$ real degrees
of freedom, while $b_{a_{1}a_{2}a_{3}a_{4}}$ has $16$ real degrees of
freedom. Thus, by this simple (degree of freedom) counting one note that it
seems more natural to associate the $4$-rebit $b_{a_{1}a_{2}a_{3}a_{4}}$
with the complex $3$-qubit, $a_{a_{1}a_{2}a_{3}}$, than with the complex $4$%
-qubit, $a_{a_{1}a_{2}a_{3}a_{4}}$. Of course, by imposing some constraints
one can always reduce the $32$ real degrees of freedom of $%
a_{a_{1}a_{2}a_{3}a_{4}}$ to $16$, and this is the kind of embedding
discussed in Ref. [9]. Here, we shall focus in the first possibility, that
is we associate the $4$-rebit $b_{a_{1}a_{2}a_{3}a_{4}}$ with the $3$-qubit $%
a_{a_{1}a_{2}a_{3}}$. The whole idea is to make sense out of a $4$-rebit in
the Nambu-Goto context without loosing the important connection with a
division algebra via the Hopf map $S^{15}\overset{S^{7}}{\longrightarrow }%
S^{8}$. Since from the point of view of division algebra the $3$-qubit is
special one may argue that $4$-rebit is also special and therefore the $%
(4+4) $-signature must also be special. Motivated by this observation one
may now proceed to recall why a straightforward application of Duff's
prescription can not be applied to the $4$-rebit. The main purpose of this
paper is to propose a solution for a connection between $4$-rebit and
Nambu-Goto action.

Before we proceed further let us add other sources of motivation concerning
the $(4+4)$-signature. First, we all agree that at macroscopic scales a
general description of our world requires $(1+3)$-dimensions (a manifold of
one time dimension and three space dimensions). But even for no experts it
is evident the lack of symmetry between the number of time and space
dimensions of our world. A natural question is: Why nature did not choose
instead of $(1+3)$-dimensions other more symmetric combinations, such as $%
(1+1)$, $(2+2)$ or $(4+4)$-dimensions? Of course, one may expect that any
complete unified theory must explain no only the number of dimensions of the
spacetime but also its signature [18]. In the lack of such a unified theory
it turns out convenient to explore separate signatures and dimensions. In
this context it has been shown that the cases $(1+1)$ and $(2+2)$ may be
considered as exceptional signatures [19]. We shall prove that in the
context of the Nambu-Goto action the target space of $(4+4)$-dimensions can
be understood as two copies of the $(2+2)$-dimensions. Roughly speaking, one
may note that this is true because $(4+4)=((2+2)+(2+2))$. Another similar
motivation can be found if one considers the combination $(4+4)=((3+1)+(1+3))
$. In other words the $(4+4)$-dimensions can be splitted in the usual $(1+3)$%
-dimensions and in $(3+1)$-dimensions. It turns out that the case $(3+1)$%
-dimensions can be considered simply as a change of signature of $(1+3)$%
-dimensions [20]. So, $(4+4)$-dimensions must contains the usual $(1+3)$%
-dimensions of our world and a mirror $(3+1)$-dimensions with the signature
changed.

Let us start by showing first that straightforward application of the Duff's
formalism concerning the Nambu-Goto action/qubits correspondence works for $%
(2+2)$-signature, but no for the $(4+4)$-signature. For the case of $(2+2)$%
-signature, consider the identification,

\begin{equation}
\begin{array}{cc}
x^{11}\leftrightarrow x^{1}+x^{3}, & x^{12}\leftrightarrow x^{2}+x^{4}, \\ 
&  \\ 
x^{21}\leftrightarrow x^{2}-x^{4}, & x^{22}\leftrightarrow -x^{1}+x^{3}.%
\end{array}
\tag{4}
\end{equation}%
Of course, this is equivalent to consider the matrix%
\begin{equation}
x^{ab}=\left( 
\begin{array}{cc}
x^{1}+x^{3} & x^{2}+x^{4} \\ 
x^{2}-x^{4} & -x^{1}+x^{3}%
\end{array}%
\right) .  \tag{5}
\end{equation}%
It is not difficult to prove that%
\begin{equation}
ds^{2}=dx^{\mu }dx^{\nu }\eta _{\mu \nu },  \tag{6}
\end{equation}%
can also be written as

\begin{equation}
ds^{2}=\frac{1}{2}dx^{ab}dx^{cd}\varepsilon _{ac}\varepsilon _{bd},  \tag{7}
\end{equation}%
where%
\begin{equation}
\eta _{\mu \nu }=diag(-1,-1,1,1),  \tag{8}
\end{equation}%
is a flat metric corresponding to $(2+2)$-signature and $\varepsilon _{ab}$
is the completely antisymmetric symbol ($\varepsilon $-symbol) with $%
\varepsilon _{12}=1$. Note that (7) is invariant under $SL(2,R)^{\otimes 2}$
transformations.

We shall now show that a generalization of (6) and (7) to a target space of $%
(4+4)$-signature leads to a line element identically equal to zero. In this
case the corresponding expressions similar to (4) are

\begin{equation}
\begin{array}{cc}
x^{111}\leftrightarrow x^{1}+x^{5}, & x^{121}\leftrightarrow x^{2}+x^{6}, \\ 
&  \\ 
x^{211}\leftrightarrow x^{2}-x^{6}, & x^{221}\leftrightarrow -x^{1}+x^{5},
\\ 
&  \\ 
x^{112}\leftrightarrow x^{3}+x^{7}, & x^{122}\leftrightarrow x^{4}+x^{8}, \\ 
&  \\ 
x^{212}\leftrightarrow x^{4}-x^{8}, & x^{222}\leftrightarrow -x^{3}+x^{7}.%
\end{array}
\tag{9}
\end{equation}%
This is equivalent to consider two matrices

\begin{equation}
x^{ab1}=\left( 
\begin{array}{cc}
x^{1}+x^{5} & x^{2}+x^{6} \\ 
x^{2}-x^{6} & -x^{1}+x^{5}%
\end{array}%
\right) ,  \tag{10}
\end{equation}%
and%
\begin{equation}
x^{ab2}=\left( 
\begin{array}{cc}
x^{3}+x^{7} & x^{4}+x^{8} \\ 
x^{4}-x^{8} & -x^{3}+x^{7}%
\end{array}%
\right) .  \tag{11}
\end{equation}

At first sight one may consider the line element

\begin{equation}
ds^{2}=\frac{1}{2}dx^{abc}dx^{def}\varepsilon _{ad}\varepsilon
_{be}\varepsilon _{cf}  \tag{12}
\end{equation}%
as the analogue of (7). But this vanishes identically because $s^{cf}\equiv
dx^{abc}dx^{def}\varepsilon _{ad}\varepsilon _{be}$ is a symmetric quantity,
while $\varepsilon _{cf}$ is antisymmetric.

Similarly, it is not difficult to show [1] (see also Ref. [8]) that the
world sheet metric in $(2+2)$-dimensions

\begin{equation}
\gamma _{ab}=\partial _{a}x^{\mu }\partial _{b}x^{\nu }\eta _{\mu \nu
}=\gamma _{ba},  \tag{13}
\end{equation}%
can also be written as%
\begin{equation}
\gamma _{ab}=\frac{1}{2}\partial _{a}x^{cd}\partial _{b}x^{ef}\varepsilon
_{ce}\varepsilon _{df}.  \tag{14}
\end{equation}%
While in $(4+4)$-dimensions, with%
\begin{equation}
\eta _{\mu \nu }=(-1,-1,-1,-1,+1,+1,+1,+1),  \tag{15}
\end{equation}%
we have%
\begin{equation}
\gamma _{ab}=\partial _{a}x^{\mu }\partial _{b}x^{\nu }\eta _{\mu \nu
}=\gamma _{ba}.  \tag{16}
\end{equation}%
But if one tries to construct the analogue of (14),

\begin{equation}
\gamma _{ab}=\frac{1}{2}\partial _{a}x^{cdg}\partial _{b}x^{fhl}\varepsilon
_{cf}\varepsilon _{dh}\varepsilon _{gl},  \tag{17}
\end{equation}%
one observes that in this case (17) implies that $\gamma _{ab}$ is
antisymmetric, that is $\gamma _{ab}=-\gamma _{ba}$, which is a
contradiction because we have in (16) that $\gamma _{ab}$ is a symmetric
matrix.

In $(2+2)$-dimensions one can write the determinant of $\gamma _{ab}$,

\begin{equation}
\det \gamma =\frac{1}{2}\varepsilon ^{ab}\varepsilon ^{cd}\gamma _{ac}\gamma
_{bd},  \tag{18}
\end{equation}%
in the form

\begin{equation}
\det \gamma =\frac{1}{2}\varepsilon ^{ab}\varepsilon ^{cd}\varepsilon
_{eg}\varepsilon _{fh}\varepsilon _{ru}\varepsilon
_{sv}b_{a}^{~~ef}b_{c}^{~~gh}b_{b}^{~~rs}b_{d}^{~~uv}=Det(b),  \tag{19}
\end{equation}%
with

\begin{equation}
b_{a}^{~~cd}\equiv \partial _{a}x^{cd}.  \tag{20}
\end{equation}%
One recognizes in (19) the hyperdeterminant of the hypermatrix $b_{a}^{~~cd}$%
. Thus, this proves that the Nambu-Goto action [2]-[3]

\begin{equation}
S=\frac{1}{2}\int d^{2}\xi \sqrt{\det \gamma },  \tag{21}
\end{equation}%
for a flat target \textquotedblleft spacetime\textquotedblright \ with $%
(2+2) $-signature can also be written as [1]

\begin{equation}
S=\frac{1}{2}\int d^{2}\xi \sqrt{Det(b)}.  \tag{22}
\end{equation}%
This process does not work for a $(4+4)$-signature because even at the level
of metric $\gamma _{ab}$ given in (16) and (17) there is a contradiction. So
for $(2+2)$ one can associate a $3$-rebit with the Nambu-Goto action, but,
by using a straightforward generalization, we have proved that this link
does not work for a target space of $(4+4)$-dimensions.

The problem with the line element $ds^{2}$ in $(4+4)$-dimensions can be
solved if instead of (12) we write

\begin{equation}
ds^{2}=\frac{1}{2}dx^{abc}dx^{def}\varepsilon _{ad}\varepsilon _{be}\eta
_{cf}.  \tag{23}
\end{equation}%
Here, we have changed the last $\varepsilon $-symbol in (12) for an $\eta $%
-symbol. But we now need to prove that (23) is equivalent to (6), with $\eta
_{\mu \nu }$ given by (15). Considering that $\eta _{cf}=diag(-1,1)$, we
find that (23) leads to%
\begin{equation}
ds^{2}=-\frac{1}{2}dx^{ab1}dx^{de1}\varepsilon _{ad}\varepsilon _{be}+\frac{1%
}{2}dx^{ab2}dx^{de2}\varepsilon _{ad}\varepsilon _{be},  \tag{24}
\end{equation}%
But each one of these terms can be identified with a space of $(2+2)$%
-signature. Therefore, one may say that the symmetry associated with (24) is
of the form $SL(2,R)^{\otimes 2}\oplus SL(2,R)^{\otimes 2}$. Indeed, we need
to redefine the matrix (10) in the form

\begin{equation}
x^{ab1}=\left( 
\begin{array}{cc}
x^{5}+x^{1} & x^{6}+x^{2} \\ 
x^{6}-x^{2} & -x^{5}+x^{1}%
\end{array}%
\right) ,  \tag{25}
\end{equation}%
while (11) remains the same. So, using (11) and (25) it is not difficult to
prove that (24) implies (6).

Similarly, the metric $\gamma _{ab}$ now becomes

\begin{equation}
\gamma _{ab}=\frac{1}{2}b_{a}^{~~cdg}b_{b}^{~~fhl}\varepsilon
_{cf}\varepsilon _{dh}\eta _{gl}.  \tag{26}
\end{equation}%
Here, the quantity $b_{a}{}^{cdg}$ is given by

\begin{equation}
b_{a}^{~~cdg}=\partial _{a}x^{cdg}.  \tag{27}
\end{equation}%
Now $\gamma _{ab}$ is symmetric in agreement with (16). Therefore, it now
makes sense to consider the determinant

\begin{equation}
\det \gamma =\frac{1}{2}\varepsilon ^{ab}\varepsilon ^{cd}\gamma _{ac}\gamma
_{bd},  \tag{28}
\end{equation}%
which implies%
\begin{equation}
\begin{array}{c}
\det \gamma =\frac{1}{2}\varepsilon ^{ab}\varepsilon ^{cd}\varepsilon
_{eg}\varepsilon _{fh}\varepsilon _{ru}\varepsilon _{sv}\eta _{pq}\eta
_{wz}b_{a}^{~~efp}b_{c}^{~~ghq}b_{b}^{~~rsw}b_{d}^{~~uvz} \\ 
\\ 
=\frac{1}{2}\varepsilon ^{ab}\varepsilon ^{cd}\varepsilon _{eg}\varepsilon
_{fh}\varepsilon _{ru}\varepsilon
_{sv}b_{a}^{~~ef1}b_{c}^{~~gh1}b_{b}^{~~rs1}b_{d}^{~~uv1} \\ 
\\ 
-\frac{1}{2}\varepsilon ^{ab}\varepsilon ^{cd}\varepsilon _{eg}\varepsilon
_{fh}\varepsilon _{ru}\varepsilon
_{sv}b_{a}^{~~ef1}b_{c}^{~~gh1}b_{b}^{~~rs2}b_{d}^{~~uv2} \\ 
\\ 
-\frac{1}{2}\varepsilon ^{ab}\varepsilon ^{cd}\varepsilon _{eg}\varepsilon
_{fh}\varepsilon _{ru}\varepsilon
_{sv}b_{a}^{~~ef2}b_{c}^{~~gh2}b_{b}^{~~rs1}b_{d}^{~~uv1} \\ 
\\ 
+\frac{1}{2}\varepsilon ^{ab}\varepsilon ^{cd}\varepsilon _{eg}\varepsilon
_{fh}\varepsilon _{ru}\varepsilon
_{sv}b_{a}^{~~ef2}b_{c}^{~~gh2}b_{b}^{~~rs2}b_{d}^{~~uv2}.%
\end{array}
\tag{29}
\end{equation}%
This can also be written as

\begin{equation}
\begin{array}{c}
\det \gamma =\frac{1}{2}(-\varepsilon
^{ab}b_{a}^{~~ef1}b_{b}^{~~rs1}+\varepsilon ^{ab}b_{a}^{~~ef2}b_{b}^{~~rs2})
\\ 
\\ 
\times (-\varepsilon ^{cd}b_{c}^{~~gh1}b_{d}^{~~uv1}+\varepsilon
^{cd}b_{c}^{~~gh2}b_{d}^{~~uv2})\varepsilon _{eg}\varepsilon
_{fh}\varepsilon _{ru}\varepsilon _{sv}.%
\end{array}
\tag{30}
\end{equation}%
Thus, introducing the variable

\begin{equation}
c^{efrs}\equiv (-\varepsilon ^{ab}b_{a}^{~~ef1}b_{b}^{~~rs1}+\varepsilon
^{ab}b_{a}^{~~ef2}b_{b}^{~~rs2}),  \tag{31}
\end{equation}%
we find

\begin{equation}
\det \gamma =\frac{1}{2}c^{efrs}c^{ghuv}\varepsilon _{eg}\varepsilon
_{fh}\varepsilon _{ru}\varepsilon _{sv}.  \tag{32}
\end{equation}%
One recognizes in (32) the hyperdeterminant of the hypermatrix $c^{efrs}$.
So, we can write

\begin{equation}
\det \gamma =Det(c).  \tag{33}
\end{equation}%
This proves that the Nambu-Goto action in $(4+4)$-dimensions%
\begin{equation}
S=\frac{1}{2}\int d^{2}\xi \sqrt{\det \gamma },  \tag{34}
\end{equation}%
can also be written as

\begin{equation}
S=\frac{1}{2}\int d^{2}\xi \sqrt{Det(c)}.  \tag{35}
\end{equation}%
Thus, we have proved that by choosing (23) instead of (12) our process also
works for a $(4+4)$-signature.

One may gain some insight on the subject if one connects qubits with the
chirotope concept in oriented matroid theory. First let us recall how this
works in a space of $(2+2)$-signature. First, one observes that (6) can be
written in the alternative Schild type [21] form

\begin{equation}
\det \gamma =\frac{1}{2}\sigma ^{\mu \nu }\sigma ^{\alpha \beta }\eta _{\mu
\alpha }\eta _{\nu \beta },  \tag{36}
\end{equation}%
where

\begin{equation}
\sigma ^{\mu \nu }=\varepsilon ^{ab}b_{a}^{\mu }b_{b}^{\nu }.  \tag{37}
\end{equation}%
Here, we have used the definition

\begin{equation}
b_{a}^{\mu }\equiv \partial _{a}x^{\mu }.  \tag{38}
\end{equation}%
It turns out that the quantity $\chi ^{\mu \nu }=sign\sigma ^{\mu \nu }$ can
be identified with a chirotope of an oriented matroid (see Refs. [22]-[24]
and also [25]-[26]). In fact, since $\sigma ^{\mu \nu }$ satisfies the
identity $\sigma ^{\mu \lbrack \nu }\sigma ^{\alpha \beta ]}\equiv 0$, one
can verify that $\chi ^{\mu \nu }$ satisfies the Grassmann-Pl\"{u}cker
relation

\begin{equation}
\chi ^{\mu \lbrack \nu }\chi ^{\alpha \beta ]}=0,  \tag{39}
\end{equation}%
and therefore $\chi ^{\mu \nu }$ is a realizable chirotope (see Ref. [22]
and references therein). Here, the bracket $[\nu \alpha \beta ]$ in (39)
means completely antisymmetric.

Since the Grassmann-Pl\"{u}cker relation (39) holds, the ground set 
\begin{equation}
E=\{ \mathbf{1,2,3,4}\}  \tag{40}
\end{equation}%
and the alternating map 
\begin{equation}
\chi ^{\mu \nu }\rightarrow \{-1,0,1\},  \tag{41}
\end{equation}%
determine a $2$-rank realizable oriented matroid $M=(E,\chi ^{\mu \nu })$.
The collection of bases for this oriented matroid is

\begin{equation}
\mathcal{B}=\{ \mathbf{\{1,2\},\{1,3\},\{1,4\},\{2,3\},\{2,4\},\{3,4\}}\}, 
\tag{42}
\end{equation}%
which can be obtained by just given values to the indices $\mu $ and $\nu $
in $\chi ^{\mu \nu }$. Actually, the pair $(E,\mathcal{B})$ determines a $2$%
-rank uniform non-oriented ordinary matroid.

In the case of qubits, the expressions (40) and (42) suggest to introduce
the underlying ground bitset (from bit and set)%
\begin{equation}
\mathcal{E}=\{1,2\}  \tag{43}
\end{equation}%
and the pre-ground set

\begin{equation}
E_{0}=\{(1,1),(1,2),(2,1),(2,2)\}.  \tag{44}
\end{equation}%
It turns out that $E_{0}$ and $E$ can be related by establishing the
identification

\begin{equation}
\begin{array}{cc}
(1,1)\leftrightarrow \mathbf{1}, & (1,2)\leftrightarrow \mathbf{2}, \\ 
&  \\ 
(2,1)\leftrightarrow \mathbf{3}, & (2,2)\leftrightarrow \mathbf{4}.%
\end{array}
\tag{45}
\end{equation}%
Observe that (45) is equivalent to making the identification of indices $%
\{a,b\} \leftrightarrow \mu $,..,etc. In fact, considering these
identifications the family of bases (42) becomes

\begin{equation}
\begin{array}{c}
\mathcal{B}_{0}=\{ \{(1,1),(1,2)\},\{(1,1),(2,1)\},\{(1,1),(2,2)\}, \\ 
\\ 
\{(1,2),(2,1)\},\{(1,2),(2,2)\},\{(2,1),(2,2)\} \}.%
\end{array}
\tag{46}
\end{equation}

Using the definition

\begin{equation}
\sigma ^{efrs}\equiv \varepsilon ^{ab}b_{a}^{~~ef}b_{b}^{~~rs},  \tag{47}
\end{equation}%
one can show that the determinant (19) can also be written as

\begin{equation}
\det \gamma =\frac{1}{2}\sigma ^{efrs}\sigma ^{ghuv}\varepsilon
_{eg}\varepsilon _{fh}\varepsilon _{ru}\varepsilon _{sv}=Det(b).  \tag{48}
\end{equation}%
This establishes a link between the hyperdeterminant (48) in terms of a
\textquotedblleft chirotope\textquotedblright \ structure (47).

If we compare (48) with (32) we see that both expressions have exactly the
same form except that $\sigma ^{efrs}$ has been replaced by $c^{efrs}$. Thus
this shows that $c^{efrs}$ can in fact be identified with a chirotope. So
one wonders whether in the case of $(4+4)$-dimensions one can go backwards
and make the identification $c^{efrs}\rightarrow c^{^{\mu \nu }}$. Let us
assume that this is possible, then we must have

\begin{equation}
c^{\mu \nu }=(-\varepsilon ^{ab}b_{a}^{\mu 1}b_{b}^{\nu 1}+\varepsilon
^{ab}b_{a}^{\mu 2}b_{b}^{\nu 2}).  \tag{49}
\end{equation}%
In turn this means that we can write

\begin{equation}
b_{a}^{\mu 1}=\partial _{a}x^{\mu }  \tag{50}
\end{equation}%
and

\begin{equation}
b_{a}^{\mu 2}=\partial _{a}y^{\mu }.  \tag{51}
\end{equation}%
Therefore, (49) becomes%
\begin{equation}
c^{\mu \nu }=(-\varepsilon ^{ab}\partial _{a}x^{\mu }\partial _{b}x^{\nu
}+\varepsilon ^{ab}\partial _{a}y^{\mu }\partial _{b}y^{\nu }).  \tag{52}
\end{equation}%
We recognize in this expression the Pl\"{u}cker coordinates for both cases $%
u_{a}^{\mu }=\partial _{a}x^{\mu }$ and $v_{a}^{\mu }=\partial _{a}y^{\mu }$.

Thus, from both quantities $\sigma ^{efrs}$ and $c^{efrs}$ (qubitopes), we
have discovered the underlying structure $Q=(\mathcal{E},E_{0},B_{0})$ which
for convenience in Ref. [8] it was called qubitoid. The word
\textquotedblleft qubitoid\textquotedblright \ is a short word for
qubit-matroid.

The above scenario can be generalized for class of $N$-qubits, with the
Hilbert space in the form $C^{2^{N}}=C^{L}\otimes C^{l}$, with $L=2^{N-n}$
and $l=2^{n}$. Such a partition allows a geometric interpretation in terms
of the complex Grassmannian variety $Gr(L,l)$ of $l$-planes in $C^{L}$ via
the Pl\"{u}cker embedding [9]. In the case of $N$-rebits one can set a $%
L\times l$ matrix variable $b_{a}^{\mu }$, $\mu =1,2,...,L$, $a=1,2...,l$,
of $2^{N}=L\times l$ associated with the variable $b_{a_{1}a_{2}...a_{N}}$,
with $a_{1},a_{2,...}etc$ taking values in the set $\{1,2\}$. In fact, one
can take the first $N-n$ terms in $b_{a_{1}a_{2}...a_{N}}$ are represented
by the index $\mu $ in $b_{a}^{\mu }$, while the remaining $n$ terms are
considered by the index $a$ in $b_{a}^{\mu }$. One of the advantage of this
construction is that the Pl\"{u}cker coordinates associated with the real
Grassmannians $b_{a}^{\mu }$ are natural invariants of the theory. Since
oriented matroid theory leads to the chirotope concept which is also defined
in terms Pl\"{u}cker coordinates these developments establishes a possible
link between chirotopes, qubitoids and $p$-branes.

In this scenario the $4$-rebit given in (31) admit a geometric
interpretation in terms of the real Grassmannian $Gr(8,2)$ or $Gr(4,4)$.
Furthermore, it may be interesting to extend to the present approach to a
line element in dimensions with $(8+8)$-signature. Apart because one can
write $(8+8)=((4+4)+(4+4))$ it is known that this kind of signatures appear
in a module space of $\mathcal{M}_{3}^{\ast }=[SO(2,2)]^{2}\backslash
SO(4,4) $, which is obtained from dimensional reduction of the moduli space $%
\mathcal{M}_{4}=[U(1)\backslash SL(2;R)]^{3}$ of the STU model of $D=4$, $%
\mathcal{N}=2$ supergravity (see Refs. [15] and [27] for details).

It is worth mentioning how could be related the present work with the Hopf
fibration $S^{15}\overset{S^{7}}{\longrightarrow }S^{8}$. This was part of
our original motivation, but we have not yet address this problem. Since it
is the complex $3$-qubit, $a_{a_{1}a_{2}a_{3}}$, which is related to such a
Hopf fibration (see Refs. [11]-[13]) our main task is to understand how the $%
4$-rebit $b_{a_{1}a_{2}a_{3}a_{4}}$ is connected with $a_{a_{1}a_{2}a_{3}}$.
The simplest (but no the most general) possibility seems to be

\begin{equation}
a_{a_{1}a_{2}a_{3}}=b_{a_{1}a_{2}a_{3}1}+ib_{a_{1}a_{2}a_{3}2}.  \tag{53}
\end{equation}%
In turn this implies

\begin{equation}
a_{a_{1}}^{~~a_{2}a_{3}}=\partial _{a_{1}}x^{a_{2}a_{3}1}+i\partial
_{a_{1}}x^{a_{2}a_{3}2}=\partial _{a_{1}}(x^{a_{2}a_{3}1}+ix^{a_{2}a_{3}2}),
\tag{54}
\end{equation}%
where we used expression (27). Therefore the 3-qubit $%
a_{a_{1}}^{~~a_{2}a_{3}}$ is related to the two $2$-rebits states $%
x^{a_{2}a_{3}1}$ and $x^{a_{2}a_{3}2}$ in the form given by (54).

Now, when one requires a normalization of the complex states $%
a_{a_{1}a_{2}a_{3}}$ the resultant space is $15$-dimensional sphere $S^{15}$
which, under the Hopf map, admit parametrization of $S^{7}$ fibration over $%
S^{8}$. It is known that $S^{7}$ is a parallelizable sphere. In fact, it has
been shown that if there exist a division algebra then the only
parallelizable spheres are $S^{1}$, $S^{3}$ and $S^{7}$ [28]-[29], which by
the Hurwitz theorem, can be associated to the complex numbers, quaternions
and octonions, respectively. Indeed, Adams [30] showed that there exist a
Hopf map $f:S^{2s-1}\longrightarrow S^{s}$ with Hopf invariant one only in $%
s=2,4$ o $8$. These remarkable results establishes the relevance of the $%
a_{a_{1}a_{2}a_{3}}$ and $S^{7}$ connection, which in turn implies a $%
a_{a_{1}a_{2}a_{3}}$ relation with octonions. It turns out, that just as the
norm group of quaternions is $SO(4)=S^{3}\times S^{3}$ , the norm group of
octonions is $SO(8)=S^{7}\times S^{7}\times G_{2}$ (see Ref. [31] and [32]
for details). Since in the $4+4$-signature the relevant group is $SO(4,4)$,
one may start asking by the $8$-dimensional spinor representation associated
with $SO(8)$. First, let us observe that $spin(8)$ admits a representation
in terms of the structure constants $(r_{i})_{j}^{k}$ of octonions $o_{i}$, (%
$o_{i}o_{j}=(r_{i})_{j}^{k}o_{k}$) namely

\begin{equation}
\left( 
\begin{array}{cc}
0 & (r_{i})_{j}^{k} \\ 
-(r_{i})_{j}^{k} & 0%
\end{array}%
\right) .  \tag{55}
\end{equation}%
Moreover, when $SO(8)$ decomposed under the subgroup $SO(4)\times SO(4)$ one
gets irreducible representation

\begin{equation}
8\longrightarrow (4,1)+(1,4).  \tag{56}
\end{equation}%
Thus, in the case of $SO(4,4)$ one may consider decomposition under the
subgroup $SO(2,2)\times SO(2,2)$ obtaining,

\begin{equation}
(4+4)\longrightarrow ((2+2),1)+(1,(2+2)).  \tag{57}
\end{equation}%
It turns out that these two direct summands correspond to the variables $%
x^{ab1}$ and $x^{ab2}$ used in (24). This explains why $dx^{abc}$, in (23),
is contracted with $\eta _{ab}$, and no with $\varepsilon _{ab}$, as in
(12). This also explains why although $dx^{abc}$ is written in three rebit
notation the invariant of (23) is $SL(2,R)^{\otimes 2}\oplus
SL(2,R)^{\otimes 2}$ rather than $SL(2,R)^{\otimes 3}$. In fact $x^{abc}$
should be understood as a two $2$-rebits rather than as a $3$-rebit. These
observations are even more evident when one considers the variables $x^{\mu }
$ and $y^{\nu }$ introduced in (50) and (51) respectively. In such cases one
has the identification $x^{ab1}\longrightarrow x^{\mu }$ and $%
x^{ab2}\longrightarrow y^{\nu }$ and consequently one may understand the
variables $x^{\mu }$ and $y^{\nu }$ as the reduction of the $(4+4)$-vector
representation of $SO(4,4)$ into the direct sum $((2+2),1)+(1,(2+2))$, given
in (57). Moreover, since considering (27) and (31) one may express $c^{abcd}$
in terms of $x^{ab1}$ and $x^{ab2}$ the expression (57) should also clarify
why $c^{abcd}$ must not be considered as a true $4$-rebit.

Finally, since through the relations (53) and (54) we have established a
possible connection between the two $2$-rebits $x^{ab1}$ and $x^{ab2}$ and
the $3$-qubit $a^{abc}$ one wonders whether the hyperdeterminant (33) or
(48) may be related to the Wong and Christensen [33] potential-entaglement
mesure $3$-tangle associated with a $3$-qubit. This link is suggested by the
fact that the analysis of the $N$-tangle formalism is different if $N$ is
even or odd. But this is precisely what we have described when one notice
that the metric in (14) associated with the variables $x^{ab}$ in $(2+2)$%
-dimensions behaves different that the metric (17) corresponding to the
variables $x^{abc}$ in $(4+4)$-dimensions. Presumably this observation may
be generalized to higher dimensions in the sense that the formalisms of $%
x^{a_{1}...a_{2s}}$ must be different that the one of $x^{a_{1}...a_{2s+1}}$%
. In this context, it may be interesting to see, for further research,
whether the analysis of the Ref. [9] of the $N$-tangle structure in terms of
the Pl\"{u}cker coordinates establishes a connection with the determinant of
the metric of the Schild type action in higher-dimensional target
`space-time'.

\bigskip \ 

\noindent \textbf{Acknowledgments: }I would like to thank the Mathematical,
Computational \& Modeling Sciences Center of the Arizona State University
where part of this work was developed. I would like also to thank an
anonymous referee for valuable comments.

\smallskip \

\end{document}